\documentclass[12pt]{article}
\usepackage{amscd,amssymb,amsmath,latexsym,enumerate}
\usepackage[mathscr]{euscript}
\usepackage{epsfig}
\usepackage{fancybox}
\usepackage{verbatim}

\usepackage{color}

\title{Topological insulators from the perspective of non-commutative geometry and index theory
}

\author{Hermann Schulz-Baldes
\\
Department Mathematik, FAU Erlangen-N\"urnberg, Germany
}

\date{ }

\newtheorem{theo}{Theorem}
\newtheorem{defini}[theo]{Definition}
\newtheorem{proposi}[theo]{Proposition}

\newcommand{\CM}{{\mathbb C}}
\newcommand{\NM}{{\mathbb N}}
\newcommand{\RM}{{\mathbb R}}

\newcommand{\TM}{{\mathbb T}}
\newcommand{\ZM}{{\mathbb Z}}
\newcommand{\PM}{{\mathbb P}}

\newcommand{\Aa}{{\cal A}}
\newcommand{\Ee}{{\cal E}}

\newcommand{\EE}{{\bf E}}

\newcommand{\Ff}{{\cal F}}

\newcommand{\Tr}{\mbox{\rm Tr}}
\newcommand{\Tt}{{\cal T}}

\newcommand{\Cc}{{\cal C}}

\newcommand{\Ll}{{\cal L}}

\newcommand{\Kk}{{\cal K}}

\newcommand{\Maj}{{\mbox{\rm\tiny Maj}}}

\newcommand{\one}{{\bf 1}}

\newcommand{\SF}{{\rm SF}} 
\newcommand{\Ch}{{\rm Ch}} 
\newcommand{\Exp}{{\rm Exp}} 
\newcommand{\Ind}{{\rm Ind}} 
\newcommand{\Ker}{{\rm Ker}} 
 
\newcommand{\Mat}{{\rm Mat}}

\newcommand{\diag}{{\rm diag}} 
\newcommand{\GFunc}{f_{\mbox{\rm\tiny Ind}}}
\newcommand{\FFunc}{f_{\mbox{\rm\tiny Exp}}}

\newcommand{\SPH}{S_{\mbox{\rm\tiny ph}}}
\newcommand{\SCH}{S_{\mbox{\rm\tiny ch}}}
\newcommand{\STR}{S_{\mbox{\rm\tiny tr}}}

\begin{document}

\maketitle

\vspace{.5cm}

\begin{abstract}
Topological insulators are solid state systems of independent electrons for which the Fermi level lies in a mobility gap, but the Fermi projection is nevertheless topologically non-trivial, namely it cannot be deformed into that of a normal insulator. This non-trivial topology is encoded in adequately defined invariants and implies the existence of surface states that are not susceptible to Anderson localization. This non-technical review reports on recent progress in the understanding of the underlying mathematical structures, with a particular focus on index theory.
\end{abstract}

\vspace{.7cm}

In solid state physics, one distinguishes between conductors and insulators. In conductors the electrons at the Fermi level can move through the solid and lead to electric currents, while in insulators there are either no such electrons due to a spectral gap or they are localized due to destructive interferences and thus cannot move freely, and one then speaks of a mobility gap due to Anderson localization.  Whether a given solid is a conductor or an insulator can be determined from a quantum mechanical one-particle Hamiltonian modeling the electrons near the Fermi level. The parameters of this so-called tight-binding model are obtained as the exchange integrals between the orbitals of the chemical compounds of the solid. 

\vspace{.2cm}

In the last decade, a multitude of both theoretical and experimental physics contributions showed that  insulators can  be topological in the sense that the Fermi projections have non-trivial topological invariants (given in terms of winding numbers, Chern numbers, etc.). These topological insulators or topological materials are distinct from conventional normal insulators, and this difference manifests itself typically by non-trivial boundary physics. More precisely, on the boundaries of a topological material there are delocalized surface modes that make the insulator conducting after all. This makes such materials also interesting for technological applications. The connection between non-trivial topology and these surface modes is called the bulk-boundary correspondence. It is a very robust principle and also the main reason that the field more recently attracted the interest of a several mathematical physicists. Indeed, methods from $K$-theory, index theory and non-commutative geometry can be put to work in the models describing topological insulators. In some situations, the analysis even allows to explain experimental facts that can be observed in a lab. This review starts out with a short description of the physics and the underlying quantum mechanical models, and then offers an (admittedly personal) overview of recent mathematical results on them. Hopefully, this allows the interested reader to get a glimpse of the intrinsic beauty due to the materialization of deep mathematical concepts in real world solid state physics systems.

\section{What is a topological insulator?}
\label{sec-WhatIs}

The title of this section and to some extent also its content may seem better fitted for a physics than a mathematics review. Indeed, what now follows are a few pages outlining a mathematicians simplistic view of the modern solid state physics background needed to describe topological insulators. The focus will here be on systems of electrons in a solid. These are microscopic particles, so there is no way of escaping quantum mechanics. On the other hand, in a first approximation, the electrons are often described as independent non-interacting particles only submitted to the Pauli exclusion principle, even though there are actually many of them in a solid and they seem to interact strongly via the Coulomb interaction. What is usually put forward as a justification is that screening plays an important role (from a distance, the charge of an electron close to a charged nucleus doesn't look so large any more) and that, anyhow, a quasiparticle description of the excitations of an interacting electron system should hold. On rigorous grounds, very little is known on how to derive the effective quantum theories used below from first principle. A further standard simplification is the so-called tight-binding approximation. This means that each atom in the solid offers the itinerant electron only a finite number of bound states, say $L$ of them. The single-electron quantum Hamiltonian $H$ in the $d$-dimensional tight-binding approximation is then a linear, self-adjoint and discrete Schr\"odinger operator on the Hilbert space $\ell^2(\ZM^d)\otimes\CM^L$ which is typically of the form
\begin{equation}
\label{eq-Hbasic}
H\;=\;
\Delta^B\,+\,W
\;.
\end{equation}
Here the operator $\Delta^B$ is the kinetic part given by a discrete magnetic Laplacian with magnetic field $B$ and $W$ is matrix-valued potential, namely an operator which commutes the position operator $X$ on $\ell^2(\ZM^d)$, trivially extended to $\ell^2(\ZM^d)\otimes\CM^L$ (for more details, see below). This potential is, unlike in scattering theory, homogeneous in space. The easiest example of such a homogeneous potential is a periodic potential, other classes are quasi-periodic and random potentials. The latter are of particular relevance for the description of solid state systems, which are rarely perfectly clean and may even be doped on purpose with impurities on purpose. There is an interesting and important physical phenomenon linked to such random potentials, and random quantum Hamiltonians in general: Anderson localization. On an intuitive level it can be described as follows. The electron wave undergoes multiple scattering processes in the random environment; all the scattering phases along the path have to be added up; this can lead to destructive interferences, namely the electron can be trapped in the random environment. This is by now known to be a very effective mechanism in dimensions $1$ and $2$ as well as for states energetically close to a band edge, and in all these cases one speaks of dynamical strong {\it Anderson localization}, and an energy interval within this regime is called a {\it mobility gap}. Many mathematical physicists have contributed to the mathematical understanding of this regime \cite{FS,AM}, but a proof for all energies in dimension $d=2$ remains a very challenging issue. On the other hand, in dimension $d=3$ and higher, physicists predict the existence of a regime of so-called weak localization in which the destructive interferences merely lead to a diffusive motion, surely slower than a ballistic spreading, but also not localized in a strict sense. To provide a rigorous proof of this is another challenging open question in mathematical physics. Now, what does all this have to do with topological insulators?

\vspace{.2cm}

As the electrons are Fermions, the zero temperature ground state for Fermi energy $\mu\in\RM$ is characterized by the Fermi projection obtained via functional calculus from the Hamiltonian:
$$
P\;=\;\chi(H\leq\mu)
\;,
$$
where $\chi$ is the indicator function. The system is then said to be an insulator if $\mu$ lies in a mobility gap of the spectrum of $H$. A special case of this is that $\mu$ lies in a true gap with no spectrum at all. Now there is something that makes the insulator topological. Let us attempt to explain this in an example, namely that of a periodic quantum Hall system. Then the spatial dimension is $d=2$ and only one state per lattice site is needed (that is, $L=1$). Furthermore the uniform magnetic flux through a unit is supposed to be rational (in units of $2\pi$). Then the Fermi projection $P$ below a gap of the spectrum can be diagonalized by a Bloch-Floquet transform $\Ff$:
$$
\Ff P\Ff^*\;=\;\int^\oplus_{\TM^2} dk\;P(k)
\;,
$$
where $\TM^2=\RM^2\slash 2\pi\ZM^2$ is the two-dimensional Brillouin torus, $P(k)$ is a projection depending smoothly on $k\in\TM^2$ and the symbol $\oplus$ indicates that we have a direct integral representation. Such a projection is, via its range, naturally identified with a vector bundle over the torus. The topological invariant is now the Chern number associated to this vector bundle. It is defined as
\begin{equation}
\label{eq-Chern2D}
\Ch_{\{1,2\}}(P)
\;=\;
\int_{\TM^2} \frac{dk}{2\pi\imath}\;\Tr\big(P(k)[\partial_{k_1}P(k),\partial_{k_2}P(k)]\big)
\;.
\end{equation}
This index $\{1,2\}$ indicates that this invariant involves derivatives in directions $1$ and $2$. Even though this is not obvious on first sight, it is not too difficult to check that $\Ch_{\{1,2\}}(P)$ is an integer which is homotopy invariant under smooth changes of $P(k)$, and, moreover, has the additivity property $\Ch_{\{1,2\}}(P+Q)=\Ch_{\{1,2\}}(P)+\Ch_{\{1,2\}}(Q)$ for two orthogonal projections $P$ and $Q$, {\it e.g.} \cite{BES}. When this integer is non-vanishing, one says that the insulator is topological. This non-trivial topology is then responsible for physical effects, in this case the quantum Hall effect, namely the Hall conductance of the system takes an integer value in units of the inverse Klitzing constant $e^2\slash h$ . The reader may have noticed that up to now, $\Ch_{\{1,2\}}(P)$ was defined merely for a periodic system for which the Bloch-Floquet transform could be applied. Hence the system is topological, but it is not a disordered system. However, the quantum Hall effect results only from the interplay between topology and Anderson localization. One of the challenges for mathematical physicists in the 1980s was to develop a theory which could define the Chern number also for disordered systems. This was achieved in the work of Bellissard \cite{Bel,Bel2} and Avron, Seiler and Simon \cite{ASS}, a situation that will be covered by the theory below. The Chern numbers are called the {\it bulk invariants} of the quantum Hall system as they are calculated merely from a model on the entire two-dimensional plane. It was also clear to physicists \cite{Hat} that these bulk systems go along with the existence of so-called chiral edge states when boundaries are introduced, {\it e.g.} by restricting the system to a half-plane and imposing Dirichlet boundary conditions. From these edge states it is possible to calculate a {\it boundary invariant} as an adequately defined winding number. Moreover, this winding number is equal to the Chern number, a fact that is now commonly referred to as the {\it bulk-boundary correspondence} (BBC). All this was put into a $K$-theoretic mathematical description in \cite{KRS} which also applies to disordered systems and will be described in some more detail below.

\vspace{.2cm}

Quantum Hall systems are historically the first topological insulators, even though this terminology was only used much latter. The breakthrough came on the theoretical side with the work of Kane and Mele \cite{KM1,KM2}. They realized that systems of independent Fermions with half-integer spin may have non-trivial topology even in presence of time-reversal symmetry (TRS). Such a symmetry means that the Fermi projection satisfies
\begin{equation}
\label{eq-TRS}
\STR^*\, \overline{P}\,\STR
\;=\;
P
\;,
\qquad
\STR\;=\;e^{-\imath \pi \sigma_2}
\;.
\end{equation}
Here $\overline{P}=\Cc P\Cc$ denotes the complex conjugate of the Fermi projection w.r.t. a given complex conjugation on Hilbert space (an anti-linear involution), and $\sigma_2=\binom{0\,-\imath}{\imath\;\;\;0}$ is the second Pauli matrix acting on the spin degrees of freedom (recall that TRS is given by complex conjugation and a rotation in spin space by $180^{\circ}$) so that $\STR$ is a real unitary squaring to minus the identity for half-integer spin, which is often called an odd TRS. Now for such Fermi projections one readily checks  that the Chern number \eqref{eq-Chern2D} vanishes. Hence the discovery \cite{KM1} that there are nevertheless two distinguishable classes of projections came as a surprise, and made solid states richer by adding the new so-called {\it quantum spin Hall phase}. It is not described here how Kane and Mele \cite{KM2} constructed for periodic systems a  $\ZM_2$-invariant distinguishing the new phase from a conventional normal insulator. However, Section~\ref{sec-symmetries} defines such an invariant as a $\ZM_2$-index of a Fredholm operator with a symmetry. The new $\ZM_2$-invariant is not as directly related to an observable quantity (as the Chern number to the Hall conductance), it is nevertheless related to a physical phenomenon, namely via the BBC to the existence of quantum mechanical boundary states. These states were, moreover, predicted to be remarkably stable under perturbations \cite{KM2}. The field then got a further boost when systems with such edge states were discovered experimentally (see the references in \cite{QZ}). In the last several years, several groups \cite{KRY,Ma} observed experimentally that these delocalized edge states are also very stable under the perturbation by very strong magnetic fields. This leads to serious doubts about the theoretical interpretation -- which after all is rooted in TRS -- and indicates that theory may not yet have reached its definite state, see Section~\ref{sec-SpinCh}. From a mathematical perspective, the aim is first to distinguish and classify Fermi projections with TRS \eqref{eq-TRS}. This is achieved using $K$-theory with symmetries which following Atiyah is called $KR$-theory. The second aim is then to calculate these invariants and to analyze the physical effects that go along with these invariants.

\vspace{.2cm}

The TRS invokes complex conjugation. It can be odd as in \eqref{eq-TRS}, or even if $\STR^2=\one$. There is another class of Fermionic systems for which the Fermi projection satisfies the similar, but distinct invariance property:
\begin{equation}
\label{eq-PHS}
\SPH^*\, \overline{P}\,\SPH
\;=\;
\one-P
\;.
\end{equation}
Here the Fermi level is $\mu=0$ and $\SPH$ is a real unitary which can either be even ($\SPH^2=\one$) or odd ($\SPH^2=-\one$). Such a system is then said to have a particle-hole symmetry (PHS) because the symmetry exchanges particles above the Fermi level with holes below the Fermi level.  A PHS naturally appears when a quadratic many-body Hamiltonian describing a superconductor is written using the Bogoliubov-de Gennes (BdG) Hamiltonian $H$ \cite{AZ}. Again such systems can be topological and may have non-trivial boundary modes, and it is interesting to distinguish the possible phases by $K$-theoretic means and index theorems. Furthermore, by combining TRS and PHS one obtains a so-called chiral symmetry (CHS) of the form
\begin{equation}
\label{eq-CHS}
\SCH^*\, {P}\,\SCH
\;=\;
\one-P
\;.
\end{equation}
Here $\SCH=\SPH\STR$, but a phase may be added to make this symmetry always even. All combinations of TRS, PHS and CHS lead to $10$ classes, the so-called Altland-Zirnbauer classes \cite{AZ} which also correspond nicely to the $10$ classical groups of Weyl if they are classified by symmetries (see {\it e.g.} \cite{AZ} or the appendix in \cite{GS}). Within each class and for every given dimension $d$ of physical space, there may or may not be different distinguishable topological phases \cite{SRFL}. A $K$-theoretic classification scheme for such phases was put forward by Kitaev \cite{Kit}, and this will be elaborated on here. One further important element making the connection to operator algebraic $K$-theory (which considers homotopy classes of both projections and unitaries) is the following. Every projection having a CHS is encoded by a unitary operator $U$ if one goes into the eigenbasis of the (even) symmetry operator $\SCH$:
\begin{equation}
\label{eq-CHS2}
P
\;=\;
\frac{1}{2}
\begin{pmatrix}
\one & U^* \\ U & \one
\end{pmatrix}
\;,
\qquad
\SCH
\;=\;
\begin{pmatrix}
\one & 0 \\ 0 & -\one
\end{pmatrix}
\;.
\end{equation}
This $U$ is also called the Fermi unitary of a chiral system and again there may be non-trivial topology encoded in it. Let us briefly discuss the Su-Schrieffer-Heeger model \cite{SSH} as a simple one-dimensional example of such a situation (a detailed discussion of this model can be found in \cite{PSB}). Its Hamiltonian acting on  $\ell^2(\ZM)\otimes\CM^2$ is of the form
$$
H
\; =\;
\tfrac{1}{2}\,S\otimes (\sigma_1+\imath \sigma_2)\; + \;\tfrac{1}{2}\,S^*\otimes (\sigma_1-\imath \sigma_2)\; +\;m \,\one\otimes\sigma_2
\;,
$$
where $S$ is the bilateral shift on $\ell^2(\ZM)$, $m\in\RM$ is a mass and $\sigma_1=\binom{0\;1}{1\;0}$ and $\sigma_2=\binom{0\,-\imath}{\imath\;\;\;0}$ as above. In the grading of the Pauli matrices,
$$
H\;=\;\begin{pmatrix} 0 & S-\imath m \\ S^*+\imath m & 0
\end{pmatrix}
\;.
$$
This Hamiltonian is off-diagonal which reflects the CHS $\SCH H \SCH=-H$ with $\SCH=\sigma_3=\binom{1\;\;\;0}{0\;-1}$. The Fermi projection $P=\chi(H\leq 0)$ hence satisfies \eqref{eq-CHS} and is of the form \eqref{eq-CHS2} as long as $m\not\in\{-1,1\}$. One finds $U=(S^*+\imath m)|S^*+\imath m|^{-1}$. Upon discrete Fourier transform, it becomes obvious that this unitary has a winding number $\Ch_{\{1\}}(U)$ equal to $1$ for $m\in(-1,1)$ and no winding for $|m|>1$. This is the robust topology of this model. If it is non-trivial it goes along with the existence of chiral zero modes at the edges, notably bound states of the half-line restriction of $H$ with energy $0$ which are also eigenvectors of $\SCH$. 

\vspace{.2cm}

Before going on, let us briefly resume what makes out a topological insulator:

\vspace{.1cm}

$\bullet$ $d$-dimensional disordered system of independent Fermions 

$\bullet$ Fermi level in a gap or a mobility gap

$\bullet$ System is submitted to a combination of basic symmetries (TRS, PHS, CHS)

$\bullet$ non-trivial topology of bulk states (winding numbers, Chern numbers, etc.)

$\bullet$ Delocalized edge modes resulting from a bulk-boundary correspondence (BBC)

\vspace{.1cm}

\noindent Let us note that there are other physical effects linked to the non-trivial topology in these systems, {\it e.g.} topological bound states at point defects. Furthermore, topological states of matter have been found in the physics literature in a great variety of other physical system: Fermions with spacial symmetries (such as reflection, rotation), interacting Fermions, bosonic systems (topological photonic crystals, topological magnons), and also in spin systems. Here we focus only on non-interacting Fermions.

\section{Short overview of the mathematical physics literature}

Following the first two papers of Kane and Mele \cite{KM1,KM2} appeared several particularly influential theoretical physics papers \cite{RSFL,QHZ,Kit,SRFL,SCR}. No attempt will be made to list further references from the abundant theoretical and experimental physics literature. There are by now several longer review papers available, a particularly nice one being \cite{QZ}. Here an overview of rigorous mathematical analysis of topological insulator systems is offered with, obviously, a focus on the results of the author and his collaborators, in particular \cite{Sch,SB}, the joint recent monograph with Emil Prodan \cite{PSB}, as well as the papers with Giuseppe De Nittis \cite{DS} and Julian Grossmann \cite{GS}. The book only covers the so-called complex classes of topological insulators, namely those Fermionic systems either have no symmetry or a chiral symmetry, but not any of the symmetries requiring a real structure (TRS and PHS). The other papers then concern the latter cases, mainly by implementing these symmetries in the complex classes. Before going into some detail in the remainder of this review, this is a good spot to at least point the reader to interesting recent contributions of several other mathematical physicists. 

\vspace{.2cm}

Several authors followed up on the physics literature by studying periodic systems. Upon Bloch-Floquet transform, one is  then naturally led to the question of classification of vector bundles. De Nittis and Gomi achieved that for several symmetry classes ({\it e.g.} in \cite{DG,DG2}), others with an approach close to homotopy theory \cite{KZ}, $K$-theory \cite{LKK} or obstruction theory \cite{FMP}. Building on \cite{Kit,SCR}, a unifying approach based on  Bott periodicity was put forward by Kennedy and Zirnbauer \cite{KZ}. A bit earlier, it was shown by Freed and Moore how twisted equivariant $K$-theory could be used for a classification of insulators \cite{FM}. This was still restricted to vector bundles, but extensions to an operator algebraic framework were given by Thiang \cite{Thi}, Kellendonk \cite{Kel} and Kubota \cite{Kub}. Focussing on quantum spin Hall systems, Graf and Porta \cite{GP} proved a bulk-boundary correspondence by functional analytic techniques, extending results from \cite{ASV}. The bulk-boundary correspondence was also approached using $T$-duality \cite{MT2} and very successfully by means of Kasparov's $KK$-theory \cite{BCR,BKR}. Katsura and Koma \cite{KK} studied the $\ZM_2$-index of a pair of projections, which is closely tied to \cite{Sch}. Other rigorous work was done by Loring and Hastings who developed local signatures to detect non-trivial invariants \cite{LH,Lor}. This has been applied to several systems now, and the connection to the more conventional index approach is currently under investigation.

\section{Bulk models and associated observable algebras}

Let us now be a bit more precise about the Fermionic one-particle bulk Hamiltonians to be studied. Here the word ``bulk" indicates that the system is extended in all directions of $d$-dimensional space and that there are no boundaries present. These operators will be of the form \eqref{eq-Hbasic} acting on a tight-binding Hilbert space $\ell^2(\ZM^d)\otimes\CM^L$, but, moreover, will be indexed by a variable $\omega$ from a compact topological space $\Omega$ modeling the disordered configurations of the solid:
\begin{equation}
\label{eq-HamDis}
H_\omega\;=\;
\Delta^B\,+\,W_\omega
\;.
\end{equation}
The discrete magnetic Laplacian for a magnetic field given by an anti-symmetric real matrix $B=(B_{i,j})_{i,j=1,\ldots,d}$ is typically of the form
$$
\Delta^B
\;=\;
\sum_{i=1}^{d} \;(t_i^*U_i+t_iU_i^*)
\;,
$$ 
where the $t_i$ are $L\times L$-matrices allowing to describe, {\it e.g.}, spin-orbit coupling, and the $U_i$ are the magnetic translations on the lattice satisfying the commutation relations
\begin{equation}
\label{eq-ComRel}
U_jU_i
\;=\;
e^{\imath B_{i,j}}U_iU_j
\;.
\end{equation}
There are many gauges encoding the magnetic translations, one being the Landau gauge presented next. Let $S_j:\ell^2(\ZM^d)\to \ell^2(\ZM^d)$ be the shift operator in the $j$th direction defined by $S_j|n\rangle=|n+e_j\rangle$ where the Dirac notation $|n\rangle=\delta_n$ for localized states on the lattice site $n\in\ZM^d$ was used and $e_j$ is the standard basis of $\ZM^d$. If then $X_j$ are the unbounded self-adjoint position operators defined by $X_j|n\rangle=n_j|n\rangle$, then the Landau gauge for $d=3$ is
\begin{equation}
\label{eq-Landau}
U_1\,=\,e^{\imath B_{1,2} X_2+ \imath B_{1,3} X_3}S_1\;,
\qquad
U_2\,=\,e^{\imath B_{2,3} X_3}S_2\;,
\qquad 
U_3\,=\,S_3
\;.
\end{equation}
This can readily be extended to higher dimensions, always such that $U_d=S_d$. The matrix potential in \eqref{eq-HamDis} is typically of the form
$$
W_\omega
\;=\;
\sum_{n\in\ZM^d}\, |n\rangle\omega_n\langle n|
\;,
$$ 
with self-adjoint $L\times L$ matrices $\omega_n$. Together these matrices form a configuration $\omega=(\omega_n)_{n\in\ZM^d}$ of the solid and the set of all such configurations is denoted by $\Omega$. All $\omega_n$ are supposed to be drawn independently and identically from a compact subset of the self-adjoint $L\times L$ matrices. Then the set $\Omega$ equipped with the Tychonov topology is also a compact space on which the product probability measure is denoted by $\PM$. For averages over $\PM$ we simple write $\EE_\PM$. Finally, there is a natural shift action $T:\ZM^d\times\Omega \to\Omega$  on $\Omega$ and $\PM$ is invariant and ergodic w.r.t. this action. It will furthermore be assumed that the support of the distribution of $\omega_n$ is a contractible set. Then also $\Omega$ is contractible.

\vspace{.2cm}

Summing up, we have a random family of Hamiltonians $H=(H_\omega)_{\omega\in\Omega}$ of the form \eqref{eq-HamDis}, indexed by points $\omega$ from a compact and ergodic dynamical system $(\Omega,T,\ZM^d,\PM)$. Actually, the detailed from of the kinetic part $\Delta_B$ and the potential part $W_\omega$ are not so important and may be generalized to include, say, random matrix elements between lattice points farther apart.  What is crucial, however, is the covariance properties of the family $(H_\omega)_{\omega\in\Omega}$, namely there are so-called dual magnetic translations $a\in\ZM^d\mapsto V_a$ commuting with the $U_j$ (and actually constructed similarly as in \eqref{eq-Landau}) so that the following covariance relation holds:
\begin{equation}
\label{eq-CovRel}
V_a H_\omega V_a^*\;=\;H_{T_a\omega}
\;,
\qquad
a\in\ZM^d
\;.
\end{equation}
Now the set of all covariant operator families of finite range (no matrix elements from $|n\rangle$ to $|m\rangle$ for $|n-m|$ arbitrarily large) forms a $*$-algebra because linear combinations and products of two such covariant families are again covariant and of finite range, and so is the adjoint. Hence one can introduce $C^*$-algebra of bulk-observables as
$$
\Aa_d
\; =\;
\mbox{\rm C}^*\,\big\{A=(A_\omega)_{\omega\in\Omega} \mbox{ finite range covariant operators}\big\}
\;.
$$
The $C^*$-closure is concretely given w.r.t. the norm $\|A\|=\sup_{\omega\in\Omega}\|A_\omega\|$ where $\|A_\omega\|$ is the operator norm on $\ell^2(\ZM^d)\otimes\CM^L$. Let us point out that if $\Omega$ consists of just one point and the magnetic field vanishes, then the covariance relation \eqref{eq-CovRel} reduces to translation invariance and, upon discrete Fourier transform, the algebra $\Aa_d$ is simply given by $C(\TM^d)\otimes\Mat(L\times L,\CM)$, the matrix valued continuous functions on the $d$-dimensional torus. In general, the algebra $\Aa_d$ can be seen to be isomorphic to a twisted crossed product algebra $C(\Omega)\rtimes_B \ZM^d$ as well as an $d$-fold iterated crossed product $C(\Omega)\rtimes \ZM\ldots \rtimes \ZM$ \cite{Bel,PSB}. Even though this is not spelled out in detail, these are important facts for the sequel. Further reasons to put the $C^*$-algebra $\Aa_d$ into the spotlight are the following:

\vspace{.1cm}

$\bullet$ $\Aa_d$ contains most physically relevant operators constructed from the Hamiltonian.

$\bullet$ $\Aa_d$ is not too large, so it still contains interesting topological information. 

$\bullet$ The $C^*$-algebraic framework allows to use the associated algebraic $K$-theory,

$\bullet$ as well as Connes' non-commutative geometry

$\bullet$ by defining a quantized calculus (non-commutative differentiation and integration).

$\bullet$ The Toeplitz extension of $\Aa_d$ connects bulk and boundary topology.

$\bullet$ Physical symmetries (TRS, PHS, CHS) can be implemented by involutive automorphisms.

\vspace{.1cm}

\noindent The aim of the remainder of this review is to support all these claims.

\section{Classification of bulk phases by $K$-theory}
\label{sec-KClass}

Topological $K$-theory classifies vector bundles over a manifold, and algebraic $K$-theory analyzes homotopy classes of projections in a C$^*$-algebra. This latter will be used to distinguish topological phases. Let us begin by explaining why, and then give a more formal definition of the $K$-groups. The bulk phase of a system of independent Fermions described by $(H_\omega)_{\omega\in\Omega}$ at zero temperature is specified by the associated Fermi projections $P_\omega=\chi(H_\omega\leq \mu)$. If $\mu$ lies in a (bulk) gap of the system, namely the Fermi level $\mu\in\RM$ is not in the spectrum $H_\omega$, then the characteristic function $\chi$ can be replaced by a continuous function, so that $P_\omega$ is obtained by continuous functional calculus from the Hamiltonian. Moreover, the family $P=(P_\omega)_{\omega\in\Omega}$ inherits the covariance property and therefore is a projection in the C$^*$-algebra $\Aa_d$. Now a continuous deformation of the Hamiltonian leads to a continuous deformation of the projection (as long as the gap remains open) and all covariant projections obtained in this manner belong to the same bulk phase of the system. Hence one is naturally led to using homotopy equivalence classes of projections in $\Aa_d$ as phase labels for the system. These classes almost define the $K_0$-group of $\Aa_d$ which also considers classes of projections in matrix algebras of $\Aa_d$:
$$
K_0(\Aa_d)
\;=\;
\{[P]_0-[Q]_0\,:\,P=P^2=P^*,\,Q=Q^2=Q^*\in\Mat(N\times N,\Aa_d) \mbox{ with }N<\infty\}
\;.
$$
Here $[\,.\,]_0$ denotes the homotopy equivalence classes, and the minus sign in the difference results from the Grothendieck construction of a group from a semigroup which is defined by $[P]_0+[P']_0=[\diag(P,P')]_0$ (see \cite{Bla0,WO,RLL} for mathematical details). From a physical point of view, the additional matrix degrees of freedom in $K$-theory reflect that there are high energy bands present which are discarded when modelling in a tight-binding framework (effectively a high energy cut-off). There seem to be no known physical effects linked to a fixed finite number of degrees of freedom per cell (the $L$ above), so that it is reasonable to classify phases by $K$-groups rather than just homotopy classes with fixed dimension of the fiber.

\vspace{.2cm}

In complex $K$-theory there is a second group $K_1(\Aa_d)$ made up of homotopy equivalence classes of unitaries, again without fixing the matrix degree of freedom:
$$
K_1(\Aa_d)
\;=\;
\{[U]_1\,:\,U=(U^*)^{-1}\in\Mat(N\times N,\Aa_d) \mbox{ with }N<\infty\}
\;.
$$
The group structure is given $[U]_1\cdot[U']_1=[\diag(U,U')]_1$,  or equivalently by $[U]_1\cdot [U']_1=[UU']_1$ in case the matrices are of same size. The group $K_1(\Aa_d)$ is used to classify $d$-dimensional Fermionic systems with a chiral symmetry. Indeed, the Fermi projection is then determined by a Fermi unitary as in \eqref{eq-CHS2}. Actually, if the chiral symmetry only holds approximately in the sense that $HJ+JH$ is sufficiently small in norm, then the block representation of $P$ as in \eqref{eq-CHS2} still has an invertible off-diagonal entry and its phase still defines a Fermi unitary and thus a class in $K_1(\Aa_d)$ \cite{PSB}. In conclusion, the two complex $K$-groups $K_0(\Aa_d)$ and $K_1(\Aa_d)$ allow to classify the Fermi projections of gapped covariant systems without further symmetry and an approximate chiral symmetry respectively. Hence it is of great interest to calculate these groups and understand their structure. The first simplifying step is elementary \cite{PSB}:

\begin{proposi}
If $\Omega$ is contractible, then $K_j(\Aa_d)=K_j(\Aa_B)$ where $\Aa_B=C^*(U_1,\ldots,U_d)$ is the rotation algebra generated by $d$ unitaries satisfying the commutation relations \eqref{eq-ComRel} associated to a given anti-symmetric real $d\times d$ matrix $B$.
\end{proposi}

As indicated above, the hypothesis of contractibility on $\Omega$ is adequate for disordered systems, however, for other models like quasicrystals it is not satisfied. The $K$-theory of the rotation algebra was first determined in a seminal paper by Pimsner and Voiculescu. Let us first state the result and then provide some further explanations. 

\begin{theo}[\cite{PV}]
$K_0(\Aa_d)=\ZM^{2^{d-1}}$ and $K_1(\Aa_d)=\ZM^{2^{d-1}}$.
\end{theo}

Let us stress that these $K$-groups are, in particular, independent of $B$ which can be viewed as a deformation parameter of the commutation relations \eqref{eq-ComRel}. For $B=0$, the rotation algebra is isomorphic to $C(\TM^d)$ as already stressed above, and its algebraic $K$-groups coincide with the topological $K$-groups classifying the vector bundles over the $d$-torus $\TM^d$. The paper by Pimsner and Voiculescu \cite{PV} also provides an algorithm to construct the generators of the $K$-groups, and it is helpful in the context of topological insulators to understand what these generators actually are in low dimensions. For $d=1$, the group $K_0(\Aa_1)=\ZM$ is generated by the identity $1$, and $K_1(\Aa_1)=\ZM$ is generated by the unitary $U_1$. For $d=2$, the group $K_0(\Aa_2)=\ZM^2$ has again $1$ as one generator, and then the Powers-Rieffel projection as the second (for $B\not=0$) or the Bott projection (for $B=0$), while $K_1(\Aa_2)=\ZM^2$ is simply generated by $U_1$ and $U_2$. For $d=3$, the group $K_0(\Aa_3)=\ZM^4$ is generated by $1$ and three Powers-Rieffel (or Bott) projection in the three coordinate planes, and $K_1(\Aa_3)=\ZM^4$ by $U_1$, $U_2$ and $U_3$, as well as one essentially new and intrinsically three-dimensional generator, which is given by the pre-image of the Powers-Rieffel projection under $K$-theoretic index map (for details on this and the following, see Chapter~4 of \cite{PSB}). This procedure can then be iterated. The generators $G_I$ of $K(\Aa_d)$ can then be labelled by subsets $I\subset\{1,\ldots,d\}$, for even cardinality of $I$ giving an element in $K_0(\Aa_d)$, and for odd in $K_1(\Aa_d)$. The set $I$ indicates which spacial directions are involved. For example, $G_{\{1,2\}}$ is the Powers-Rieffel projection and  $G_{\{1,2,3\}}$ the intrinsically three-dimensional generator described above. The class of a projection $P\in \Mat(N\times N,\Aa_d)$ can be decomposed as
\begin{equation}
\label{eq-Pdecomp}
[P]_0
\;=\!
\sum_{I\subset\{1,\ldots,d\},\,|I|\,{\rm even}}
\!n_I\;[G_I]_0\;,
\qquad
n_I\in\ZM
\;.
\end{equation}
The coefficients $n_I$ are called the topological invariants of $P$ (up to a constant) and the aim below will be to calculate them by non-commutative geometry (actually rather non-commutative differential topology). For $I={\{1,\ldots,d\}}$, the integer $n_I$ is called the strong topological invariant (or also top invariant), all other invariants are called weak. The reason for this terminology is that the top invariants are known to remain stable in the Anderson localization regime \cite{PLB,PSB}. It is believed that the weak invariants are less robust, but this issue is still under investigation.  This concludes the discussion of the so-called complex classes of topological insulators, namely those not invoking a real structure (needed to implement TRS and PHS). The same approach can also be used in these cases, and it will be briefly explained in Section~\ref{sec-symmetries} how to proceed.

\section{Topological invariants and their main properties}
\label{sec-BulkInv}

Let us begin by introducing the so-called non-commutative analysis tools on the algebra $\Aa_d$. Integration is given by tracial normalized state on $\Aa_d$ specified by the trace per unit volume:
$$
\Tt(A)
\;=\;
\lim_{N\to\infty}\frac{1}{(2N+1)^d}\,\sum_{n\in [-N,N]^d}\,\langle n|A_\omega|n\rangle
\;=\;
\EE_{\mathbb{P}}\; 
\langle 0| A_\omega|0\rangle
\;,
\qquad
A=(A_\omega)_{\omega\in\Omega}\in\Aa_d
\;.
$$
The convergence for $N\to\infty$ holds $\PM$-almost surely by Birkhoff's ergodic theorem (for the ergodic action of $\ZM$ on $\Omega$), which, due to the covariance relation, also shows that the almost sure value is given by the average over $\PM$ on the r.h.s.. Non-commutative derivations $\nabla_j$ are densely defined by
$$
\nabla_{j}A_\omega
\;=\;
\imath[X_j,A_\omega]
\;,
\qquad
j=1,\ldots,d
$$
where the position operators are given by $X_j|n\rangle=n_j|n\rangle$. Operators in $\Aa_d$ which are infinitely many times differentiable are called smooth, and these smooth operators form a Fr\'echet sub-algebra of $\Aa_d$. Both $\Tt$ and $\nabla_j$ naturally extend to matrix algebras $\Mat(N\times N,\Aa_d)$ over $\Aa_d$. One has the invariance property $\Tt(\nabla_j A)=0$ so that partial integration holds. Furthermore, if one deals with a periodic system so that $\Aa_d=C(\TM^d)$,  the trace $\Tt$ reduces to the integration of the Brillouin torus $\TM_d$ and $\nabla_j$ to the partial derivative w.r.t. the $j$-th component of the quasimomentum. Hence the non-commuative analysis tools naturally extend the well-known semiclassical analysis approach to solid state physics systems. Equipped with these tools, there is now a standard procedure to produce Connes-Chern characters on $\Aa_d$ \cite{Con} spelled out next.  The somewhat lengthy formulas cannot be avoided, but are very natural when the full structure behind them is uncovered (as in \cite{Con}), and there are even good reasons for the choice of the normalization constants, as will become apparent further below.

\begin{defini}
\label{def-CC}
For even cardinality $|I|$ and a smooth projection $P\in \Mat(N\times N,\Aa_d)$, the $I$-th Chern number of $P$ is
$$
\mbox{\rm Ch}_{I} (P)
\;=\;
\frac{(2\imath \pi)^{\frac{|I|}{2}}}{\frac{|I|}{2}!}\;  \sum_{\rho\in S_I} (-1)^\rho \,\Tt\left(P\prod_{j=1}^{|I|} \nabla_{\rho_j} P  \right)
\;.
$$
Here the sum runs over bijections $\rho:\{1,\ldots,|I|\}\to I$ for which the definition of the signature $(-1)^\rho$ is extended using the natural order on $I$. For odd $|I|$ and an invertible smooth $A\in\Mat(N\times N,\Aa_d)$, the $I$-th Chern number is
$$
\mbox{\rm Ch}_I (A)
\;=\;
\frac{\imath(\imath \pi)^\frac{|I|-1}{2}}{|I|!!}\;  \sum_{\rho\in S_{I}} (-1)^\rho \;\Tt
\left(\prod_{j=1}^{|I|}A^{-1} \nabla_{\rho_j}A  \right)
\;,
$$
where $(2n+1)!!=\prod^n_{k=1}(2k+1)$.
\end{defini}

Before going on to describe the mathematical properties of these objects, let us stress that they often linked to physical quantities. It is by now a classic fact that $\mbox{\rm Ch}_{\{1,2\}} (P)$ is the zero-temperature Hall conductance of two-dimensional electron system described by the Fermi projection $P$ \cite{BES}. Actually, for a periodic system $\mbox{\rm Ch}_{\{1,2\}} (P)$ reduces to \eqref{eq-Chern2D} after Fourier transform. Furthermore, if one of the $d$ direction is interpreted as time, it is essentially the orbital polarization of a periodically driven system \cite{ST,DL} and then $\mbox{\rm Ch}_{\{1,2,3,4\}} (P)$ describes the non-linear magneto-electric response \cite{PLB,PSB}. For chiral systems with Fermi unitary $U$, $\mbox{\rm Ch}_{\{j\}}(U)$ is the so-called chiral polarization \cite{PSB}. Further examples are described in \cite{PSB}. All these quantities are of topological nature due to the following classic result.

\begin{theo}[\cite{Con}]
$\mbox{\rm Ch}_{I} (P)$ and $\mbox{\rm Ch}_I (A)$ are homotopy invariants under smooth deformations of $P\in\Aa_d$ and  $A\in\Aa_d$ and thus establish pairings with $K_*(\Aa_d)$.
\end{theo}

The rather complicated choice of the normalization constants in Definition~\ref{def-CC} guarantees, beneath other things, that the top pairings with $I=\{1,\ldots,d\}$ are integer valued, and that the generalized Streda formulas described below hold. Furthermore, by a suspension argument \cite{PSB} a connection to the invariants introduced by Essin and Gurarie \cite{EG1} (based on earlier work by Volovik) can be established. They  express the invariants in terms of the resolvents and hence fairly directly in terms of the Hamiltonian.

\begin{theo}[Link to Volovik-Essin-Gurarie invariants, \cite{PSB}] 
Let $|I|$ be even. Consider a path $z:[0,1]\to \CM\setminus\sigma(H)$ encircling $(-\infty,\mu]\cap\sigma(H)$ and set $G(t)=(H-z(t))^{-1}$. Then for the Fermi projection $P=\chi(H\leq \mu)$ of the gapped Hamiltonian $H$,
$$
\mbox{\rm Ch}_{I}(P_\mu)
\;=\;
\frac{(\imath \pi)^\frac{|I|}{2}}{\imath (|I|-1)!!} \sum_{\rho\in S_{I\cup \{0\}}}\!\! (-1)^\rho\! \int^{1}_0 \!dt\,\Tt\!
\left(\prod_{j=1}^{|I|}G(t)^{-1} \nabla_{\rho_j}G(t)  \right)
\;,
$$
where $\nabla_0=\partial_t$ and $\rho:\{1,\ldots,|I|+1\}\to \{0\}\cup I$. A similar formula holds for odd $|I|$.
\end{theo}

The next result concerns the dependence of the invariants on the magnetic field. Interestingly, the magnetic derivatives of weak invariants are given by other invariants. The first such relation was found by Streda \cite{Str} who considered two-dimensional quantum Hall systems and showed that the integrated density of states grows linearly in the magnetic field with derivative essentially given by the Chern number. The integrated density of states is actually $\EE\, \langle 0|P|0\rangle=\mbox{\rm Ch}_\emptyset(P)$, and Streda's formula then states
$$
\partial_{B_{1,2}}\, \mbox{\rm Ch}_\emptyset(P)
\;=\;
\frac{1}{2\pi}\;\mbox{\rm Ch}_{\{1,2\}}(P)
\;.
$$
In this generality, the Streda formula was first proved by Rammal and Bellissard \cite{RB}, and their techniques can be extended to obtain the following. 

\begin{theo}[Generalized Streda formul\ae, \cite{PSB}]
\label{theo-Streda}
For sufficiently smooth projections $P$ and invertibles $A$ and $|I|$ even and odd respectively, one has for $i,j\not\in I$
$$
\partial_{B_{i,j}}\,\mbox{\rm Ch}_{I}(P)\;=\;\frac{1}{2\pi}\;
\mbox{\rm Ch}_{I\cup \{i,j\}}(P) 
\;,
\qquad
\partial_{B_{i,j}}\,\mbox{\rm Ch}_{I}(A)
\;=\;\frac{1}{2\pi}\;
\mbox{\rm Ch}_{I\cup \{i,j\}}(A) 
\;.
$$
\end{theo}

Apart from the quantum Hall effect, this result is relevant for magneto-electric effects in dimension $d=3$. Here time is added as a $4$th direction needed for calculation of polarization, and the non-linear response is the derivative w.r.t. the magnetic field of the polarization, which is thus connected to an integer $\mbox{\rm Ch}_{\{1,2,3,4\}}(P)$. Another example in $d=3$ is the chiral magneto-electric response \cite{PSB}. Furthermore, combining Theorem~\ref{theo-Streda} with results by Elliott \cite{Ell} leads to the following statement on the ranges of parings.

\begin{theo}[\cite{Ell,PSB}]
\label{theo-PairingRange}
Let $I,J\subset\{1,\ldots,d\}$ be increasingly ordered sets with a cardinality of equal parity. Then, with the generators $G_J$ of $K(\Aa_{d})$ described above, 
\begin{align}
\label{eq-paringrangeeven1}
& \Ch_{I}(G_J) 
\;=\;
0\;,
& I\setminus J\not=\emptyset\;,
\\
\label{eq-paringrangeeven2}
& \Ch_{I}(G_J) 
\;=\;
1 \;, & I=J\;,
\\
\label{eq-paringrangeeven3}
& \Ch_{I}(G_J) 
\;=\;
(2\pi)^{-\frac{1}{2}| J\setminus I|} \; {\rm Pf}(B_{J\setminus I})\;,
& I\subset J\;,
\end{align}
where $ {\rm Pf}$ denotes the Pfaffian and $B_{J\setminus I}$ is the antisymmetric matrix retaining just the indices of $J$ not contained in $I$.
\end{theo}

When these identities are combined  with \eqref{eq-Pdecomp} and the additivity property
$$
\Ch_I(P)
\;=\!
\sum_{I\subset\{1,\ldots,d\},\,|I|\,{\rm even}}
\!n_I\;\Ch_I(G_I)
\;,
\qquad
n_I\in\ZM
\;,
$$
the whole range of $\Ch_I(P)\in\RM$ is indeed determined with its full dependence on the magnetic field. Let us note that, in particular, the strong topological invariants for $I=\{1,\ldots,d\}$ are integer valued, as are the next lower weak ones with $|I|=d-1$.  Actually, for the strong topological invariants (also of a chiral system), an index theorem is known. This is described next. Let $\gamma_1,\ldots,\gamma_d$ be an irreducible representation of the complex Clifford algebra $\CM_d$ on $\CM^{2^{(d-1)/2}}$. Then the Dirac operator on $\ell^2(\ZM^d)\otimes\CM^L\otimes \CM^{2^{(d-1)/2}}$ is defined as 
\begin{equation}
\label{eq-Dirac}
D\;=\;\sum_{j=1}^d X_j\otimes \one\otimes\gamma_j
\;.
\end{equation}
This may not look like a Dirac operator at first sight, but after Fourier transform it does become a first order differential operator on the Brillouin torus. The Dirac phase is then defined by $F=D|D|^{-1}$ (after having lifted the non-relevant kernel of $D$). It can be checked that $F$ defines a so-called odd Fredholm module over $\Aa_d$, namely (by definition) $F^2=\one$ and the commutators $[F,A_\omega]$ are compact for all $A=(A_\omega)_{\omega\in\Omega}\in\Aa_d$. Actually for smooth $A$ the commutators $[F,A_\omega]$ even lie in the Schatten ideal $\Ll^{d+\epsilon}$. For even $d$, one furthermore has a grading induced by $\Gamma=-i^{-d/2}\gamma_1\cdots \gamma_d$ which anti-commutes with $D$. After an adequate basis change on the representation space $\CM^{2^{(d-1)/2}}$, $\Gamma=\diag(\one,-\one)$ and then $F=\binom{0\;\;G}{G^*\;0}$. An odd Fredholm module over $\Aa_d$ together with such a grading defines an even Fredholm module, so that is what one has for even $d$. By a standard procedure \cite{Con0,Con} one now obtains Fredholm operators associated to these Fredholm modules. Non-standard  is, however, the connection to the invariants defined in Definition~\ref{def-CC}. Such a connection is called a local index formula. Possibly it follows, similar as in \cite{Andr}, from the general Connes-Moscovici local index formula \cite{CGX}, but the proofs in \cite{PLB,PSB} are based on interesting new geometric identities for the volume of higher-dimensional simplices, which generalize Connes' two-dimensional triangle identity \cite{Con0,ASS,BES}.

\begin{theo} [Local index formula, \cite{PLB,PSB}]
\label{theo-IndexPair}
For even $d$ and a smooth projection $P\in\Aa_d$, the operator $P_\omega GP_\omega$ is $\PM$-almost surely Fredholm with almost sure index given by
$$
\Ind(P_\omega GP_\omega)
\;=\;
\mbox{\rm Ch}_{\{1,\ldots,d\}} (P)
\;.
$$
For odd $d$, let $E=\frac{1}{2}(F+\one)$ be the Hardy projection for $F$. For invertible smooth $A\in\Aa_d$, the operator $EA_\omega E$ is $\PM$-almost surely Fredholm with almost sure index given by
$$
\mbox{\rm Ind}(E\,A_\omega E)
\;=\;
\mbox{\rm Ch}_{\{1,\ldots,d\}} (A)
\;.
$$
\end{theo}

For $d=2$, $G=\frac{X_1+\imath X_2}{|X_1+\imath X_2|}$ and the theorem provides the well-known connection between the Hall conductance given by the Kubo formula and a Fredholm index \cite{BES,ASS}.
One substantial advantage of the index theorem is that it can be extended to the regime of strong Anderson localization \cite{PLB,PSB}, for which the Fermi level lies in a mobility gap. Here the Fermi projection is {\it not } in the C$^*$-algebra, but only in the subspace of sufficiently smooth elements of the enveloping von Neumann algebra with $\Tt(|\nabla P|^p)<\infty$ for $p>d$ so that $\Ch_{\{1,\ldots,d\}} (P)$ is well-defined (in the terminology of \cite{BES}, this is the defining property of a non-commutative Sobolev space which has, however, {\it not} the property of lying in the C$^*$-algebra as in the commutative case). Hence one can use the strong invariants and the associated indices as a phase label also in this localization regime. This is of crucial importance, for example, for the explanation of the quantum Hall effect \cite{BES}. 

\vspace{.2cm}

As a final comment on bulk topological invariants let us mention the approach by Loring and Hastings \cite{LH,Lor} which introduced local signature indices for topological insulators which have the advantage that they can be very easily calculated numerically for a given model and even be used to define local changes in the ground states. While this is far from obvious, it is now becoming evident that these invariants coincide with the above (work in preparation with Terry Loring).

\section{Bulk-boundary correspondence}
\label{sec-BBC}

Up to now, only bulk topological invariants were introduced and used to distinguish bulk ground states of systems of independent Fermions. One of the key features of topological insulators is that these very invariants are responsible for a large number of effects linked to defects. Most prominent is the existence of surface, edge or boundary states. These states appear in the bulk gap for the models restricted to half-spaces. That wave equations on half-spaces have such modes was already known to Rayleigh and Sommerfeld, however, the boundary states in topological insulators, moreover, come with a remarkable stability. They are not susceptible to Anderson localization if perturbed by a weak random potential (which is particularly surprising if the boundary is one-dimensional) and appear for any type of boundary condition, as long as it is local (for example, the spectral boundary conditions of Atiyah-Patodi-Singer are non-local). Less prominent, but just as important and based on similar mathematical principles, are bound states attached to particular types of point defects in topological insulators and this will be briefly described in Section~\ref{sec-SF}.

\vspace{.2cm}

Let us begin by describing the half-space Hamiltonian, for sake of concreteness by simply imposing Dirichlet boundary conditions.  We choose $\ZM^{d-1}\times\NM\subset\ZM^d$ as the half-space, with the restriction $x_d\geq 0$ on the last coordinate. Let $\Pi$ be the projection from $\ell^2(\ZM^d)$ to $\ell^2(\ZM^{d-1}\times\NM)$, naturally extended to $\ell^2(\ZM^d)\otimes\CM^L$. Then the half-space Hamiltonians acting on $\ell^2(\ZM^{d-1}\times\NM)\otimes\CM^L$ are simply given by $\widehat{H}_\omega=\Pi H_\omega \Pi$. This operator as well as all other half-space operators will carry a hat. It lies in the C$^*$-algebra $\widehat{\Aa}_d$ of half-space operators which is generated by $\widehat{U}_j=\Pi U_j\Pi$ and the functions $f\in C(\Omega)$ acting as a potential $f_\omega|n\rangle=f(T_n\omega)|n\rangle$, $n\in\ZM^{d-1}\times\NM$. Note that the $\widehat{U}_j$ are still unitaries for $j=1,\ldots,d-1$, but $\widehat{U}_d$ is now only a partial isometry. In the Landau gauge \eqref{eq-Landau} it is actually simply the unilateral shift so that $\widehat{U}_d\widehat{U}_d^*=\one-\Pi_d$ where $\Pi_d$ is the projection on the boundary surface. The C$^*$-algebra generated by $\Pi_d$, $\widehat{U}_1,\ldots,\widehat{U}_{d-1}$ and the $f_\omega$ is called the boundary or edge algebra and denoted by $\Ee_d$. It is naturally embedded in $\widehat{\Aa}_d$, which in turn is naturally mapped onto $\Aa_d$ by discarding the hats and all summands containing $\Pi_d$. Hence

\begin{proposi}[\cite{KRS,PSB}]
The boundary, half-space and bulk algebra from a short exact sequence of C$^*$-algebras
\begin{equation}
\label{eq-ExactSeq}
\begin{array}{ccccccccc}
0 & \to & \Ee_d & \to  & \widehat{\Aa}_d & \to &\Aa_d & \to & 0\;.
\end{array}
\end{equation}
Moreover, there is an isomorphism $\Ee_d\cong\Aa_{d-1}\otimes\Kk(\ell^2(\NM))$. 
\end{proposi}

Actually, \eqref{eq-ExactSeq} is isomorphic to the well-known Pimsner-Voiculescu short exact sequence for the crossed product $\Aa_d=\Aa_{d-1}\rtimes\ZM$. Associated to any short exact sequence of C$^*$-algebras there is the exact sequence of $K$-theory:
$$
\begin{array}{cccccccccc}
& K_0(\Ee_d) & {\longrightarrow} & & K_0(\widehat{\Aa}_d) & & {\longrightarrow}& & K_0(\Aa_d)&
\\
&  & & & & & & & &
\\
& \Ind\;\;\uparrow  & & & & & & & \downarrow\;\;\mbox{\rm Exp}&
\\
&  & & & & & & & &
\\
& K_1(\Aa_d)  & {\longleftarrow} & & K_1(\widehat{\Aa}_d) & & {\longleftarrow} & & K_1(\Ee_d) &
\end{array}
$$
By stability of $K$-theory and the above proposition, one has $K_j(\Ee_d)\cong K_j(\Aa_{d-1})$. Pimsner and Voiculescu, moreover, proved $K_j(\widehat{\Aa}_d)\cong K_j({\Aa}_{d-1})$ and that the maps from $K_j(\Ee_d) $ to $K_j(\widehat{\Aa}_d)$ are trivial. The definition of the interesting connecting maps (the exponential map Exp and the index map Ind) will not be described in detail here (see \cite{Bla0,WO}), but rather only the outcome for topological insulators. This is the $K$-theoretic content of the BBC.

\begin{theo}[\cite{KRS,PSB}] 
\label{theo-KBBC} Let the open set $\Delta\subset\RM$ be a gap of $H\in\Aa_d$ and $P=\chi(H\leq\mu)$ be the Fermi projection to $\mu\in\Delta$. Then
\begin{equation}
\label{ExpMapFormula}
\mbox{\rm Exp}([P]_0)
\;=\;
\left[\exp(2\pi\imath\, \FFunc(\widehat{H}))\right]_1
\;,
\end{equation}
where $\FFunc :\RM\to [0,1]$ is a non-decreasing continuous function equal to $0$ below $\Delta$ and to $1$ above $\Delta$. If $H$ has a chiral symmetry and $U$ is the Fermi unitary given by \eqref{eq-CHS2}, then
\begin{equation}
\label{IndMapFormula}
\mbox{\rm Ind} \big ([U]_1 \big )
\;=\;
\left[
e^{-\imath\frac{\pi}{2} \GFunc (\widehat{H})}
\,
{\rm diag}(\one_L,0_L)
\,e^{\imath\frac{\pi}{2} \GFunc (\widehat{H})}
\right]_0
\;-\;
\left[{\rm diag}(\one_L, 0_L)
\right]_0
\;,
\end{equation}
where $\GFunc :\RM\to [-1,1]$ is a non-decreasing smooth odd function, equal to $\pm 1$ above/below $\Delta$. If, moreover, there is a gap in the surface spectrum and $\widehat{P}=\widehat{P}_++\widehat{P}_-$ is the decomposition of the projection $\widehat{P}$ on the central surface band into sectors of positive and negative chirality, then 
$$
\mbox{\rm Ind} \big ([U]_1 \big )\;=\;[\widehat{P}_+]_0-[\widehat{P}_-]_0
\;.
$$
\end{theo}

It is remarkable that the r.h.s. in \eqref{ExpMapFormula} and \eqref{IndMapFormula} indeed define classes in $\Ee_d$, even though they are obtained by functional calculus of $\widehat{H}$ which lies in $\widehat{\Aa}_d$ and not in $\Ee_d$. Hence Theorem~\ref{theo-KBBC} connects $K$-theoretic invariants of the bulk and the boundary. 

\vspace{.2cm}

We have seen in Section~\ref{sec-BulkInv} how to extract numerical Chern-Connes invariants from the bulk classes. The same can be done for the boundary invariants. Indeed, Definition~\ref{def-CC} verbatim transposes to define pairings with projections and unitaries in $\Ee_d\cong\Aa_{d-1}\otimes\Kk(\ell^2(\NM))$. One merely has to include a trace over $\Kk(\ell^2(\NM))$ and consequently verify a supplementary trace class condition, which is satisfied, for example, for the r.h.s. of  \eqref{ExpMapFormula} and \eqref{IndMapFormula}. Again the Connes-Chern invariants over $\Ee_d$ are denoted by $\Ch_I$, but $I$ now cannot contain the index $d$. The connection to the bulk invariants is as follows.

\begin{theo}[\cite{KRS,PSB}]
\label{theo-BBC}
Provided the traceclass conditions hold, one has for $I\subset\{1,\ldots,d-1\}$ of even and odd cardinality respectively,
\begin{equation}
\label{eq-BBC}
\mbox{\rm Ch}_{I\cup \{d\}}(U)\;=\;\mbox{\rm Ch}_I(\Ind(U)) 
\;,
\qquad 
\mbox{\rm Ch}_{I\cup \{d\}}(P)\;=\;\mbox{\rm Ch}_I(\mbox{\rm Exp}(P))  
\;.
\end{equation}
Here $\Ind(U)$ and $\mbox{\rm Exp}(P)$ are representatives of $\Ind([U]_1)\in K_0(\Aa_d)$ and $\mbox{\rm Exp}([P]_0)\in K_1(\Aa_d)$.
\end{theo}

The proof in \cite{KRS,PSB} is an explicit calculation based on a formula for the index map and the suspension. Recently, a $KK$-theoretic proof has been put forward \cite{BCR,BKR} which, however, uses the full arsenal of Kasparov's $KK$-theory. This is a far reaching generalization of $K$-theory. The $K$-groups are described by generalized Fredholm operators on Hilbert modules, similar in spirit to the Atiyah-J\"anich theorem. Furthermore, every exact sequence of C$^*$-algebras, such as \eqref{eq-ExactSeq},  also leads to a $KK$-group element. The key element of the theory is then the Kasparov product of two generalized Fredholm operators describing $KK$-group elements. We will not attempt to describe any details on all of this (see \cite{Bla0}), but rather sketch for the reader familiar with $KK$-theory how it is applied in the present context. Let us consider the second identity in \eqref{eq-BBC} more precisely. The key fact is that the exact sequence \eqref{eq-ExactSeq} defines an extension class and thus an element $[\xi]\in KK^1(\Aa_d,\Ee_d)$. Its Kasparov product $[P]_0\times [\xi]\in KK^0(\CM,\Aa_d)$ with $[P]_0\in K_0(\Aa_d)\cong KK^1(\CM,\Ee_d)$ is $[U]_1$, if $[U]_1=\Exp([P]_0)$. On the other hand, using $K$-homology $[\Ch_{I}]\in KK^0(\Ee_d,\CM)$ and $[\Ch_{I\cup \{d\}}]\in KK^0(\Aa_d,\CM)$, and the Kasparov product can be shown to yield $[\xi]\times [\Ch_{I}]=[\Ch_{I\cup \{d\}}]$. Hence by the associativity of the Kasparov product, $[U]_1\times [\Ch_{I}]=[P]_0\times [\xi]\times [\Ch_{I}]=[P]_0\times [\Ch_{I\cup \{d\}}]$, but this is just claim. Yet another novel approach to the BBC is based on $T$-duality \cite{MT2}. Let us exhibit two concrete instances to which the general Theorem~\ref{theo-BBC} about the duality of pairings associated to the exact sequence \eqref{eq-ExactSeq} applies.

\begin{theo}[Quantization of boundary currents, \cite{KRS}]
\label{theo-BoundCur}
Let $d=2$ and $\mu$ lie in a bulk gap. Then the Hall conductance $\mbox{\rm Ch}_{\{1,2\}}(P)$ as given by the Kubo formula dictates the quantization of chiral boundary currents:
$$
\mbox{\rm Ch}_{\{1,2\}}(P)
\;=\;
\mbox{\rm Ch}_{\{1\}}(\mbox{\rm Exp}(P))
\;=\;
\EE_\PM \sum_{n_2\geq 0} \langle 0,n_2| \FFunc'(\widehat{H})\imath[X_1,\widehat{H}]|0,n_2\rangle
\;.
$$
\end{theo}

The first equality here is just restating Theorem~\ref{theo-BBC}, and the second one results from a calculation. The r.h.s. is physically interpreted as the quantum mechanical expectation value of the current flowing along the boundary, measured by the current operator $\imath[X_1,\widehat{H}]$, of all states in the density matrix $\FFunc'(\widehat{H})$ of the half-space operator. This density matrix can be used to approximate $|\Delta|^{-1}\chi_\Delta(\widehat{H})$ and the expression appearing in Theorem~\ref{theo-BoundCur} can be interpreted as the net contribution of boundary currents (left edge minus right edge due to different chemical potentials). The fact that this difference is quantized with the same integer as the bulk conductivity is crucial for the existence of the integer quantum Hall effect. The second application concerns a surface quantum Hall effect in chiral systems, which is termed {\it anomalous} because no external magnetic field is needed. The surface Hall conductance $\Ch_{\{1,2\}}(\Ind(U))$ is rather dictated by the bulk topological invariant. No chiral insulator where this can be measured seems to be known.

\begin{theo}[Anomalous surface quantum Hall effect, \cite{PSB}]
\label{theo-AnoSur}
Let $d=3$ and $\mu$ lie in a bulk gap. Suppose the system is chiral and described by a Fermi unitary $U$, and that the surface spectrum has a gap, opened {\it e.g.} by a weak magnetic field perpendicular to surface. Then the last claim of Theorem~\ref{theo-KBBC} applies and Theorem~\ref{theo-BBC} shows
$$ 
\mbox{\rm Ch}_{\{1,2,3\}}(U)
\;=\;
\mbox{\rm Ch}_{\{1,2\}}(\widehat{P}_+)-\mbox{\rm Ch}_{\{1,2\}}(\widehat{P}_-)
\;.
$$ 
\end{theo}

It is conjectured that generically either $\widehat{P}_+=0$ or $\widehat{P}_-=0$. In this situation, a non-trivial bulk-invariant $ \mbox{\rm Ch}_{\{1,2,3\}}(A)\not =0$ leads by Theorem~\ref{theo-AnoSur} to a surface quantum Hall effect. As a last result let us state what can indeed be proved about the delocalized nature of the surface spectrum. 

\begin{theo}[\cite{PSB}]
Suppose that $d$ is even and the Fermi projection below the bulk gap has a non-trivial strong invariant $\mbox{\rm Ch}_{\{1,\ldots,d\}} (P)\not=0$. Then for a disordered potential in an arbitrary finite strip along the boundary, the surface spectrum is not Anderson localized in the sense the Aizenman-Molcanov estimate \cite{AM,DDS} on low moments of the Green matrix does not hold for any energy in the bulk gap. For odd $d\geq 3$ and a Fermi unitary with  $\mbox{\rm Ch}_{\{1,\ldots,d\}} (U)\not=0$, there is no Anderson localization of the surface states at $\mu=0$.
\end{theo}

\section{Discrete real symmetries and periodic table}
\label{sec-symmetries}

Except for the second part of Section~\ref{sec-WhatIs}, only systems without symmetry or with a chiral symmetry have been considered so far. The ground states of such systems were classified by complex $K$-theory. Now time reversal symmetry (TRS) and particle hole symmetry (PHS) will be considered, both of which require a real structure allowing to define the complex conjugate of an operator. The ground state (Fermi projection $P$) of systems having these symmetries satisfy respectively 
\begin{equation}
\label{eq-STRPHS}
\STR^*\,\overline{P}\,\STR\;=\;P\;,
\qquad
\SPH^*\,\overline{P}\,\SPH\;=\;\one\,-\,P
\;.
\end{equation}
Both symmetry operators $\STR$ and $\SPH$ are real and square to either $\one$ or $-\one$, so that one can either have an even or and odd TRS/PHS. Furthermore, these two operators are strictly local and act in a translation invariant manner only on the fiber $\CM^L$ of the Hilbert space. It ought to be stressed again that the PHS is actually not a physical symmetry of the system, but rather a property of the BdG Hamiltonian describing a quadratic Fermionic Hamiltonian on Fock space, see {\it e.g.} \cite{KZ}. As already pointed out in Section~\ref{sec-WhatIs}, whenever both TRS and PHS are present, they combine to a chiral symmetry. Hence there are $8$ real symmetry classes, listed in the lower part of Table~\ref{table2}. Within each so-called Cartan-Altland-Zirnbauer (CAZ) symmetry class and depending on the dimension $d$ of physical space, one can now distinguish possible ground states. Again this can be done using $K$-theory by studying homotopy classes of projections satisfying the corresponding symmetry relations \eqref{eq-STRPHS} for the given CAZ class. This is done by means of so-called $KR$-theory \cite{Kar,BL}. If there is no magnetic field and the system is periodic, one is then lead to use the group $KR_{j}(C(\TM^d_\tau))$ where $j$ is as indicated in Table~\ref{table2} and $\tau$ denotes the involution $\tau(k)=-k$ on $\TM^d$. This group can be calculated \cite{Kar,Kit} and contains all the strong and weak invariants. It is rather large (like its complex counterpart described in Section~\ref{sec-KClass}), and also contains so-called torsion invariants which are $\ZM_2$-valued. To get a more clear representation of the possible phases, in a first step one restricts the attention to the strong invariants, which are known to persist in the Anderson localization regime, as will be described below. The strong invariants are those which are intrinsically $d$-dimensional and thus stem from the groups $KR_{j}(C_0(\RM^d_\tau))$. Here $\tau$ still is the $k$-space inversion. These groups are well-known \cite{Kar} to coincide with the homotopy groups of the stable orthogonal group $O$:
\begin{equation}
\label{eq-invarform}
KR_{j}(C_0(\RM^d_\tau))
\;=\;
\pi_{j-1-d}(O)
\;,
\end{equation}
given by
\begin{equation}
\label{tab-HomO}
\begin{tabular}{|c||c|c|c|c|c|c|c|c|}
\hline
$j$ & $0$ & $1$ & $2$ & $3$ & $4$ & $5$ & $6$ & $7$  \\
\hline
$\pi_{j}(O)$  &$\ZM_2$ & $\ZM_2$ & $0$ & $2\,\ZM$ & $0$ & $0$ & $0$ & $\ZM$ 
\\
\hline
\end{tabular}
\;.
\end{equation}

\begin{table}
\begin{center}
\begin{tabular}{|c||c|c|c||c||c|c|c|c|c|c|c|c|}
\hline
$\!\!j\backslash d\!\!$ &$\!$TRS$\!$ & $\!$PHS$\!$ & $\!$CHS$\!$ & $\!$CAZ$\!$ &  $1$  &  $2$ & $3$ & $4$ & $5$ &  $6$ & $7$ & $8$ 
\\\hline\hline
$0$ & $0$ &$0$&$0$ & A &  & $\ZM$ & & $\ZM$ & & $\ZM$  & & $\ZM$  
\\
$1$ & $0$&$0$&$ 1$ & AIII  &$\ZM$ & & $\ZM$ & & $\ZM$ & & $\ZM$  & 
\\
\hline\hline
$0$ & $+1$&$0$&$0$ & AI & & & & $2\,\ZM$ & & $\ZM_2$ & $\ZM_2$  & $\ZM$
\\
$1$ & $+1$&$+1$&$1$ & BDI & $\ZM$ & & & & $2\,\ZM$ & & $\ZM_2$ & $\ZM_2$  
\\
$2$ & $0$ &$+1$&$0$ & D & $\ZM_2$ & $\ZM$ & & & & $2\,\ZM$ & & $\ZM_2$  
\\
$3$ & $-1$&$+1$&$1$  & DIII & $\ZM_2$ & $\ZM_2$ & $\ZM$ & & & & $2\,\ZM$ & 
\\
$4$ & $-1$&$0$&$0$  & AII & & $\ZM_2$ & $\ZM_2$ & $\ZM$ & & & & $2\,\ZM$
\\
$5$ & $-1$&$-1$&$1$  & CII & $2\,\ZM$ & & $\ZM_2$ & $\ZM_2$ & $\ZM$ & & & 
\\
$6$ & $0$ &$-1$&$0$  & C & & $2\,\ZM$ & & $\ZM_2$ & $\ZM_2$ & $\ZM$  & & 
\\
$7$ & $+1$&$-1$&$1$  & CI  & & & $2\,\ZM$ & & $\ZM_2$ & $\ZM_2$  &  $\ZM$ & 
\\
\hline
\end{tabular}
\caption{\sl So-called periodic table of topological insulators, listing possible values of the strong invariants for the CAZ classes in dependence of the dimension $d$.
}
\label{table2}
\end{center}
\end{table}

\vspace{.1cm}

\noindent Bott periodicity states that $\pi_j(O)=\pi_{j+8}(O)$. Reporting this data into a table now produces the celebrated periodic table for topological insulators \cite{Kit,RSFL}. Why there appears a $2\,\ZM$ instead of the isomorphic group $\ZM$ will be explained below.  The above discussion did not explain why the CAZ classes are placed in the particular order labelled by the index $j$. Indeed, it requires quite a lengthy series of arguments of successively implementing symmetries to show that the CAZ classes correspond to the index $j$ as stated, see \cite{SCR,FM,Thi,KZ,GS,Kel}. 

\vspace{.2cm}

An alternative approach \cite{Sch,GS} is to directly implement the symmetries in the index pairings described in Theorem~\ref{theo-IndexPair}. This has the advantage that disordered systems can be dealt with even when the Fermi energy lies in a mobility gap. Let first explain how this works by considering the example of a two-dimensional system having an odd TRS (just as in the first novel topological insulator studied in \cite{KM2}). Then $\STR^* \overline{P}\STR=P$ with $\STR^2=-\one$.  As the Dirac phase $G=\frac{X_1+\imath X_2}{|X_1+\imath X_2|}$ commutes with $\STR$ and satisfies $G^t=G$ where $A^t=(\overline{A})^*$ denotes the transpose of an operator $A$. The Fredholm operator $T_\omega=P_\omega G P_\omega$ in Theorem~\ref{theo-IndexPair} thus satisfies $\STR^* (T_\omega)^t\STR=T_\omega$, or equivalently the Fredholm operator $T_\omega\STR$ is anti-symmetric. It can then be proved \cite{AS,Sch} that the set of anti-symmetric Fredholm operators has exactly $2$ connected components labelled by the compactly stable and homotopy invariant $\ZM_2$-valued index
\begin{equation}
\label{eq-Ind2}
\mbox{\rm Ind}_2(T)
\;=\;
\dim(\mbox{\rm Ker}(T))\;\mbox{\rm mod}\;2
\;\in\;\ZM_2
\;.
\end{equation}
As already stressed above, this allows to define a  $\ZM_2$ phase label for Kane-Mele model in the regime of a mobility gap. As a second example, let us consider a system with odd TRS in dimension $d=8$. Then one can check that the Dirac phase satisfies $\overline{G}=G$. Consequently, the Fredholm operator $T_\omega=P_\omega G P_\omega$  satisfies $\STR^* \overline{T_\omega}\STR=T_\omega$. It is hence quaternionic and has an even dimensional kernel and cokernel so that the Fredholm index is even (thus the entry $2\,\ZM$ for $(j,d)=(4,8)$ in Table~\ref{table2}). This parity can actually be of some physical significance, see Theorem~\ref{theo-PHSbound} below.

\vspace{.2cm}

Let us now sketch how this approach can be extended to treat all cases appearing in the periodic table. The symmetry of the Fermi projection (even/odd TRS/PHS) is one ingredient entering into the symmetry of the index pairings of Theorem~\ref{theo-IndexPair}, a second one comes from the symmetry of the homological part given by the Dirac operator \eqref{eq-Dirac}. In the above, this was $G^t=G$ and $\overline{G}=G$, and the general situation is described next. The irreducible representation $\gamma_1,\ldots,\gamma_d$ of the Clifford algebra $\CM_d$ is chosen such that $\gamma_{2j}=-\overline{\gamma_{2j}}$ and $\gamma_{2j+1}=\overline{\gamma_{2j+1}}$. Recall that for $d$ even there   is a grading $\Gamma=-i^{-d/2}\gamma_1\cdots \gamma_d$.  Now one can show that there exists a real unitary $\Sigma$ (essentially unique) such that
\begin{center}
\begin{tabular}{|c|c|c|c|c|c|c|c|c|}
\hline
$d=8-i$ & $8$ & $7$ & $6$ &  $5$ &  $4$ &  $3$ &  $2$ &  $1$ 
\\
\hline
${\Sigma}^2$ & $\one$ & $\one$ & $-\one$ & $-\one$ & $-\one$ & $-\one$ & $\one$ & $\one$ 
\\
\hline
${\Sigma}^*\,\overline{D}\,{\Sigma}$ & $D$ & $-D$ & $D$ & $D$ & $D$ & $-D$ & $D$ & $D$ 
\\
\hline
$\Gamma\,{\Sigma}\,\Gamma$ & ${\Sigma}$ & & $-{\Sigma}$ & & ${\Sigma}$ & & $-{\Sigma}$ & 
\\
\hline
\end{tabular}
\;.
\end{center}
The construction of this representation is explicit \cite{GS} and is related to irreducible representations of the real Clifford algebra. The data of the table sates that $(D,\Gamma,\Sigma)$ defines a $KR^i$-cycle, also called a spectral triple with real structure (this is particularly clear by comparing with \cite{GVF}). Now one can do a careful symmetry analysis of the index pairings in Theorem~\ref{theo-IndexPair}. Even though this can be achieved with basic functional analytic tools, there are actually a number of intricate spectral degeneracy arguments needed (one of them being the classical Kramers' degeneracy).

\begin{theo}[\cite{GS}] 
The strong index paring from Theorem~\ref{theo-IndexPair} leads to a $\ZM$, $\ZM_2$ or $2\,\ZM$ index as stated in Table~\ref{table2}, provided the Fermi level lies in a region of Anderson localization.
\end{theo}

As there are index theorems for the boundary invariants in complete analogy with Theorem~\ref{theo-IndexPair}, the very same techniques also allows to define $\ZM$, $\ZM_2$ or $2\,\ZM$ indices for the boundary invariants and then establish a BBC for these invariants based on the implementation of symmetries in Theorem~\ref{theo-KBBC}, but there is no published report on this.  As there is a very intimate relation between Bott periodicity and real Clifford algebras, it is also possible to prove Clifford-module-valued index pairings \cite{LM} for the strong bulk invariants in the realm of $KK$-theory \cite{BCR2} (this is still restricted to gapped systems for now). This also allows to connect bulk and boundary invariants in a natural manner, by an extension of the argument sketched after Theorem~\ref{theo-BBC}, see \cite{BKR}. Once all these invariants are defined, an important question is about their physical significance. Some further insight is presented in the remaining two sections.

\vspace{.3cm}

\section{Spectral flow in topological insulators}
\label{sec-SF}

The spectral flow of a given family $t\in[0,1]\mapsto T_t$ of self-adjoint Fredholm operators is the net number of eigenvalues moving through $0$, namely those moving up count for $1$ and those moving down with $-1$ (if simple, otherwise the multiplicity appears). For analytic paths, one can readily refer to analytic perturbation theory in order to make a rigorous definition out of this basic idea, but for merely continuous paths this is somewhat delicate and was only made precise by Phillips \cite{Phi}. The basic properties of the spectral flow are its homotopy invariance (with fixed endpoints) and a natural concatenation property. Furthermore, there is a close link to index theory for paths of with unitarily equivalent end points $T_1=U^*T_0U$, namely their spectral flow is equal to the index of the Fredholm operator $PUP$ where $P=\chi(T_0)$ \cite{Phi}. In topological insulators, the parameter $t$ stems from a local perturbation of the Hamiltonian denoted by $H_t$, and then $T_t=H_t-\mu$ for $\mu$ in a gap of $H=H_0$ is a family of self-adjoint Fredholm operators. The main example of the type is the insertion of a flux tube in a two-dimensional system, that is the addition of a supplementary magnetic field $2\pi t$ through one cell of the lattice $\ZM^2$. This situation and actually also the result below are usually referred to as a Laughlin argument, see for example \cite{BES,ASS}. While the gauges used to realized the (local) flux tube are not compact, one can check that $T_t=H_t-\mu$ is nevertheless a path of Fredholm operators \cite{Mac,DS}. 

\begin{theo}[\cite{Mac,DS}]
Let $d=2$ and $\mu$ lie in a gap of $H=H_0$. Set $P=\chi(H\leq\mu)$. The spectral flow associated to the insertion of a flux tube satisfies
$$
\mbox{\rm SF}(t\in[0,1]\mapsto H_t-\mu)
\;=\;
\Ch_{\{1,2\}}(P)
\;.
$$
\end{theo}

It is now possible to analyze the fate of this spectral flow in a two-dimensional system with odd TRS, namely in a quantum spin Hall system. As the Chern number $\Ch_{\{1,2\}}(P)$ vanishes, there is actually no spectral flow. However, the path has the symmetry property $\overline{H_t}=-\STR^* G^* H_{1-t} G\STR$ with $G$ as in Theorem~\ref{theo-IndexPair}. Therefore also the spectral curves have symmetry property $\sigma(H_t)=-\sigma(H_{1-t})$. Let us further point out that the insertion of the flux breaks the odd TRS, except for a half-flux $t=\frac{1}{2}$ which can be interpreted as a local defect respecting the symmetry. Thus one has Kramers' degeneracy for eigenvalues of $H_0$, $H_{\frac{1}{2}}$ and $H_1$. It can be shown \cite{ASV} that there are two topologically distinct classes of spectral curves having these Kramers' degeneracies and the symmetry described above. The non-trivial one appears exactly when the bulk $\ZM_2$ invariant is non-trivial. Thus one can conclude that the half-flux defect necessarily leads to a Kramers' degenerate bound state, if the system is topologically non-trivial:

\begin{theo}[\cite{DS}] Let $d=2$ and suppose that $\mu$ is in a gap of $H=H_0$ having odd TRS. If the $\ZM_2$-index $\Ind_2(P_\omega GP_\omega)$, defined as in \eqref{eq-Ind2}, is non-trivial, then $H_{\frac{1}{2}}$ has  Kramers degenerate bound state in the gap of $\mu$.
\end{theo}

It is also possible to consider other symmetry classes. Particularly interesting are the BdG Hamiltonians with PHS. Insertion of a flux breaks the PHS, but again a half-flux is local defect respecting PHS. Now the attached bound state is a so-called Majorana zero mode (namely in second quantization, it induces a self-adjoint creation operator).

\begin{theo}[\cite{DS}] 
\label{theo-PHSbound}
Let $d=2$ and suppose that $0$ is in a gap of $H=H_0$ having an even or odd PHS. Then
$$
\dim (\Ker(H_{\frac{1}{2}}))
\;\mbox{\rm mod }2
\;=\;
\Ch_{\{1,2\}}(P)
\;\mbox{\rm mod }2
\;.
$$
Here the Chern number $\Ch_{\{1,2\}}(P)$ of $P=\chi(H\leq 0)$ is the strong invariant of the system. If it is odd, $H_{\frac{1}{2}}$ must have a zero mode. For a system with odd PHS, $\Ch_{\{1,2\}}(P)$ is even and thus Majorana zero modes are unstable in the sense that they can be lifted by a generic perturbation.
\end{theo}

Let us briefly describe another type of spectral flow which is of interest in the context of topological insulators, as recently shown in \cite{CPS}. Consider again a BdG Hamiltonian $H$ with an even PHS $\SPH^*H\SPH=-H$. Suppose $\SPH=\binom{0 \;1}{1\;0}$ and let $C=2^{-\frac{1}{2}}\binom{1\,-\imath}{1\;\;\imath}$ be the Cayley transform in the same grading. Then the so-called Majorana representation $H_\Maj=C^* HC$ is purely imaginary and can thus be written as $H_\Maj=\imath T$ with a skew-adjoint real operator $T$, which is Fredholm provided that $0$ lies in a gap of $H$. Therefore paths of gapped BdG operators lead to paths of skew-adjoint Fredholm operators. For such paths one can define a $\ZM_2$-valued spectral flow $\SF_2$ by counting the number orientation changes of the eigenfunctions at eigenvalue crossings through $0$ \cite{CPS}. A particular example of this is obtained by inserting a flux into the one-dimensional Kitaev chain with even PHS and even TRS given by
\begin{equation}
\label{eq-KitaevDef}
H
\;=\;
\frac{1}{2}\,
\begin{pmatrix}
S+S^*+2\mu & \imath(S-S^*) \\
\imath(S-S^*) & -(S+S^*+2\mu)
\end{pmatrix}
\;+\;
W
\;,
\end{equation}
where $W$ a random matrix potential respecting the  even PHS and the even TRS $\STR^*\overline{W}\STR=W$ with $\STR=\binom{1 \;\,0}{0\;-1}$. The operator acts on $\ell^2(\ZM)\otimes\CM^2=\ell^2(\ZM\times\{0,1\})$ and $S$ is the bilateral shift as before. Now one can again insert a flux tube trough one cell of the lattice strip $\ZM\times\{0,1\}$ which now leads to a path $H_t$ of BdG operators and thus also a path of skew-adjoint Fredholm operators. If $|\mu|<1-\|W\|$, this path can be shown to have non-trivial $\ZM_2$-valued spectral flow and this non-triviality can again be used to show the existence of bound states to the defect given by a half-flux.

\begin{theo}[\cite{CPS}]
\label{theo-defect}
For $|\mu|<1-\|W\|$, the Kitaev Hamiltonian ${H}_{\frac{1}{2}}$ with a half-flux defect has an odd number of evenly degenerate zero eigenvalues:
$$
\tfrac{1}{2}\,\dim(\Ker({H}_{\frac{1}{2}}))\;\mbox{\rm mod}\,2
\;=\;1
\;.
$$
\end{theo}

There are several other situations where the spectral flow is interesting in topological insulators \cite{PSB,CPS}. What is, however, missing is a spectral flow in chiral three-dimensional systems which detects their strong topological invariant. 

\section{Spin Chern numbers}
\label{sec-SpinCh}

The last sections documented that the $\ZM_2$ invariants are well-established on a theoretical side and can be used to distinguish phases of systems having exact symmetries (like TRS and PHS). On the other hand, their importance for experimental observations is still disputed.  In several recent experiments on (two-dimensional) quantum spin Hall systems \cite{KRY,Ma}, it became apparent that the effect of having delocalized surface modes is remarkably stable under perturbation by magnetic fields (up to $9$ Tesla!). As these magnetic fields break TRS, there is no notion of $\ZM_2$-invariant in this situation and the question is whether there is another mechanism leading to these stable surface states. Here it will be argued that the so-called spin Chern numbers introduced by Prodan \cite{Pro} are topological invariants that may be at the root of the phenomena. Their definition does not require TRS, but rather an approximate conservation of one component of the spin, say $s^z$. In mathematical terms, this is expressed by requiring that
\begin{equation}
\label{eq-SCon}
\|[H,s^z]\|\;\leq\;C
\;,
\end{equation}
for some sufficiently small constant $C$. If then $P$ is the Fermi projection below the gap, one can show that the spectrum of the self-adjoint operator $Ps^zP$ has a gap at $0$ (the origin itself lies in the spectrum, but is irrelevant here). Therefore there are two Riesz projections $P_\pm$ associated to the positive and negative spectrum and they provide an orthogonal decomposition $P=P_++P_-$. Furthermore, the existence of the gap readily implies that both $P_\pm$ are smooth so that the Chern numbers $\Ch_{\{1,2\}}(P_\pm)$ are well-defined. One has the sum rule $\Ch_{\{1,2\}}(P_+)+\Ch_{\{1,2\}}(P_-)=\Ch_{\{1,2\}}(P)$ and, as typically $\Ch_{\{1,2\}}(P)=0$, it is sufficient to consider one of these invariants. The spin Chen number is then defined as $\mbox{\rm SCh}(P)=\Ch_{\{1,2\}}(P_+)$. It clearly also exists for systems without TRS. Moreover, it is connected to the $\ZM_2$-invariants discussed above.

\begin{theo}[\cite{Sch}]
Let $H$ have odd TRS and approximate spin conservation \eqref{eq-SCon}. Then
$$
\mbox{\rm Ind}_2(P_\omega G P_\omega)
\;=\;
\mbox{\rm SCh}(P)\;\mbox{\rm mod}\;2
\;.
$$ 
\end{theo}

While this result was stated for $d=2$ and odd TRS, it can readily be extended to other situations where there is an approximately conserved observable with spectral gap. Hence in many situations, there may be non-trivial invariants resulting from complex pairings present in systems with non-trivial $\ZM_2$ invariants. This is similar to the situation of invariants in chiral systems, which remain stable also when the system is only approximately chiral. It still remains to discuss the importance of spin Chern number for physical effects. A partial answer is provided by the following result which should be compared with Theorem~\ref{theo-BoundCur}.

\begin{theo}[\cite{SB}]
If  $\mbox{\rm SCh}(P)\not =0$, then spin filtered edge currents in of states in the bulk gap are stable w.r.t. perturbations by magnetic field and disorder:
$$
\EE_\PM \sum_{n_2\geq 0} \langle 0,n_2| \FFunc'(\widehat{H})\,\tfrac{1}{2}\big\{\imath[X_1,\widehat{H}],s^z\big\}|0,n_2\rangle
\;=\;
\mbox{\rm SCh}(P) \;+\; \mbox{corrections}
\;, 
$$
where the corrections depend linearly on the size of the commutator \eqref{eq-SCon} and the $C^6$-norm of $\FFunc$. Here $\{\,,\,\}$ denotes the anti-commutator used to produce a selfadjoint observable.
\end{theo}

The presence of corrections here implies that the l.h.s. is not quantized, but it is non-vanishing when the spin Chern number is non-vanishing. What is still somewhat unsatisfactory about this result is that the spin current itself is ill-defined when $[H,s^z]\not=0$. Nevertheless, the result clearly hints at the existence of surface states with continuous spectrum, even when TRS is broken.

\section{Future directions} 

Topological materials still remain a very active field of research in physics. Those currently investigated include topological photonic crystals, topological bosonic systems, topological mechanical and optomechanical systems, spin systems, as well as the role of interactions in Fermionic systems. As the long list of references suggests, there is also a growing interest in the mathematical physics community. Let us list a few open questions within the framework of topological insulators made up of independent Fermions: 

\vspace{.1cm}

$\bullet$ Index theory for weak topological invariants and definite answer on their stability; 

$\bullet$  more detailed description of the bulk-edge correspondence in cases with real symmetries;

$\bullet$  topological invariants linked to spacial symmetries (like rotation, inversion, reflection);

$\bullet$   an analysis of the stability of the invariants in the previous item; 

$\bullet$  further investigation of physical implications of the invariants, {\it e.g.} for heat transport. 

\vspace{.1cm}

\noindent Some of these issues should be settled in the near future. Also applications to the other physical systems listed above should be within reach for rigorous analysis. A major challenge, however, remains the definition and analysis of topological invariants in interacting systems and spin systems - even though there are already numerous contributions from theoretical physicists.

\vspace{.5cm}

\noindent {\bf Acknowledgements:} The author thanks his main collaborators Jean Bellissard, Johannes Kellendonk and Emil Prodan as well as further coauthors Giuseppe De Nittis,  Stefan Teufel, Julian Grossmann,  Carlos Villegas, Julio Cesar Avila, Alan Carey and John Phillips for inspiring intellectual input and persistence. This work is in part supported by the DFG. 


\end{document}